# Holistic View of the Road Transportation System Based on Real-time Data Sharing Mechanism

Tao Li[1]; Xiang Dong[1]; Junfeng Hao[2]; PingYin[1]; Xiaoxue Xu[1]; Maokai Lai[1]; Yuan Li[1]; and Ting Peng[1*]

[1*] Key Laboratory for Special Area Highway Engineering of Ministry of Education, Chang'an University, Xi'an 710064, China.
[2] China Railway No.7 Bureau Group No.3 Engineering Co.,Ltd. Xi'an 710032, China.

Corresponding Author(s). E-Mail: t.peng@ieee.org
Contributing authors:lt1186166770@163.com; 1290740787@qq.com; 1609972221@qq.com; 935524208@qq.com; 2112534668@qq.com; 3412303785@qq.com; 157708198@qq.com

**ABSTRACT**

Traditional manual driving and single-vehicle-based intelligent driving have limitations in real-time and accurate acquisition of the current driving status and intentions of surrounding vehicles, leading to vehicles typically maintaining appropriate safe distances from each other. Yet, accidents still frequently occur, especially in merging areas; meanwhile, it is difficult to comprehensively obtain the conditions of road infrastructure. These limitations not only restrict the further improvement of road capacity but also result in irreparable losses of life and property. To overcome this bottleneck, this paper constructs a space-time global view of the road traffic system based on a real-time sharing mechanism, enabling both road users and managers to timely access the driving intentions of nearby vehicles and the real-time status of road infrastructure.

**Keywords：** Road Section Management Unit; Vehicle Intelligent Units; Information sharing; Spatio-temporal global view.

## INTRODUCTION

The traffic safety situation is increasingly severe, with the number of casualties from traffic accidents increasing year by year. According to the Global Road Safety Report 2023, an estimated 1.19 million people died in road traffic accidents in 2021, equivalent to 15 deaths per 100,000 people. According to the latest in-depth analysis of traffic accidents released by the Ministry of Transport of China, the main reasons for the frequent occurrence of traffic accidents include fatigue driving, adverse weather conditions, negligence of driving environment, lack of awareness of following distance and vehicle condition, as well as other significant factors such as drunk driving, vehicle malfunctions, and speeding. These factors vary in their contribution to traffic accidents but all pose serious threats to traffic safety. Adverse weather conditions, inadequate awareness of following distance and vehicle condition, and insufficient consideration of driving environment are the main reasons, accounting for 41% of total traffic accidents. Essentially, these accidents stem from traditional manual driving and single-vehicle intelligent driving cannot accurately and timely obtain the driving intentions of other vehicles and the condition of road infrastructure.

In order to solve the above problems, this paper proposes a spatial and temporal global view of the road transportation system based on a real-time information sharing mechanism. The real-time information sharing mechanism makes it possible for road users and managers to obtain the real-time status information of vehicles and road infrastructure in the neighboring areas as well as the future driving intentions. With the help of spatio-temporal global view, comprehensive monitoring and real-time analysis of road infrastructure operation can be realized, so that more scientific and reasonable traffic management measures can be formulated.

## THE BASIC FRAMEWORK

### The Road Section Management Unit

The Road Side Unit (RSU) is deployed on both sides of the road to support cooperative intelligent transportation system services and effectively detect and collect information about passing vehicles and road anomalies. It then stores, processes, and disseminates this information.

In this paper, a novel design of a road section unit is proposed and named the Road Section Management Unit (RSMU). This innovative device not only inherits the core functions of traditional road section units but also undergoes a series of functional expansions and optimizations to meet the increasingly complex requirements of intelligent traffic systems. Figure 1. provides a clearer illustration of the functions of the Road Section Management Unit.

(1)Real-time sharing of all vehicles' driving intentions: The RSMU shares the driving intentions uploaded by Vehicle Intelligent Units in real-time with other vehicles within the jurisdiction area. All vehicles can obtain the latest driving intentions and road information through sharing, enabling drivers to better predict road conditions and make corresponding driving decisions.

(2)Processing and management of road condition information: The RSMU is connected to intelligent monitoring devices. Front-end sensors collect road condition information, and the data is transmitted to the RSMU using 4G/5G technology for sharing. It analyzes and predicts the data to draw conclusions, and shares real-time road conditions with Vehicle Intelligent Units.

(3)Abnormal condition handling function: High-definition cameras are installed on the RSMU to monitor traffic conditions within the management scope. In the event of traffic accidents, rock falls, bridge fractures, and other obstacles affecting vehicle passage, it aggregates the collected road surface information and shares it in real-time with Vehicle Intelligent Units, allowing vehicles to take timely detours or evasive actions to prevent subsequent accidents.

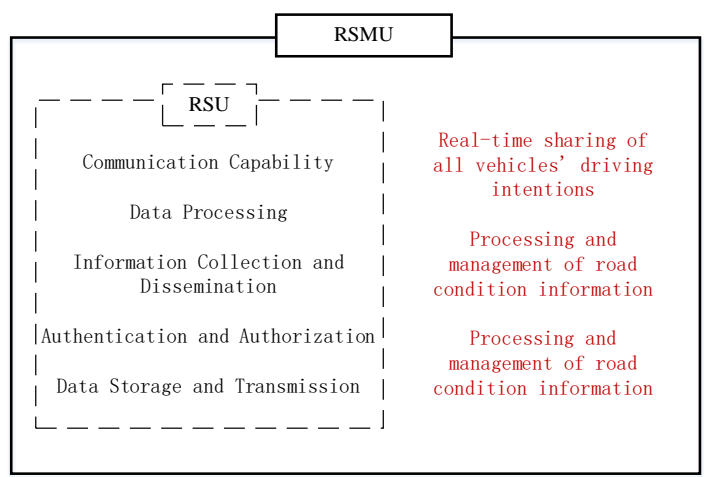

**Figure 1. Diagram of road section management unit functionality**

**Functional requirements of Vehicle Intelligent Units**

Vehicle Intelligent Units (VIU), as a device specially designed for vehicles, plays a core role in facilitating seamless communication between vehicles and other vehicles as well as road section infrastructure, thereby achieving efficient collaboration and intelligent management during the driving process. Typically, the VIU consists of GPS modules, communication modules, processors, storage units, antennas, etc., which are responsible for obtaining vehicle location information, communicating with road section facilities, processing data, storing map data, and receiving and transmitting signals. The VIU in this paper should have the following functions:

(1)Vehicle Driving Data Collection: The VIU should have modules equipped with high-performance gyroscopes, steering angle sensors, acceleration sensors, and cameras installed on the vehicle to capture relevant information about the vehicle's driving status. This information includes speed, direction, acceleration, braking status, and steering angle.

(2)Information Sharing: The VIU should transmit the real-time driving status and driving intentions to the road section management unit promptly. Additionally, it should receive information packets from the road section management unit, allowing it to obtain real-time data on the driving status of other vehicles and the status of road infrastructure. This mechanism strengthens safety communication and collaboration between vehicles and enables real-time warning functionality for potential dangers, ensuring vehicles can adjust flexibly according to environmental conditions.

(3)Driving Environment Perception: The VIU should be capable of perceiving the driving environment to ensure safe vehicle operation. Its purpose is to provide comprehensive and accurate environmental data to vehicles, thereby providing a solid safety guarantee for the driving process. To achieve this, the system needs to comprehensively collect and analyze various driving environment information, including dynamic data on surrounding vehicles, the position of pedestrians, and non-motorized vehicles.

**Implementation**

Install shortwave time signal receivers in VIU and RSMU. These precision receivers with excellent reception performance can accurately receive the standard time signals from the National Time Service Center's shortwave time signal stations. The signals serve as precise and stable time references for VIU and RSMU. Thus, they can receive the time signals in real-time and convert them into system-recognizable time data via internal decoding mechanisms.

This design ensures that VIU and RSMU can maintain accurate time synchronization, thus avoiding information errors and confusion caused by time asynchrony. Whether in complex traffic environments or high-speed moving VIU, these receivers can operate stably, ensuring the accuracy of time.

**Sharing mechanism**

In the current field of vehicle-road cooperative systems communication, there are mainly two standard technologies. One is the Dedicated Short-Range Communication (DSRC) technology led by the United States, which has wide application and recognition in the industry. The other is the Cellular Vehicle-to-Everything (C-V2X) communication technology led by China, which has shown strong development potential and application prospects domestically and globally. In fact, DSRC and C-V2X have significant differences in their working principles.

The DSRC system is composed of two key elements, the VIU and the RSU. Their cooperation facilitates bidirectional information transfer among vehicles and between vehicles and road infrastructure. Then, the RSU utilizes optical fibers or mobile networks as communication means to accurately and efficiently convey this vital traffic information to the backend intelligent transportation system platform, ensuring the smooth flow and coordinated operation of information in the transportation network. In contrast, the C-V2X system mainly encompasses the VIU, RSU, Uu interface, and PC5 interface. The RSU is primarily responsible for broadcasting details of road conditions, signal light status, and dynamic pedestrian movements within its coverage, along with providing time and position synchronization. Additionally, it has mobile network access capabilities to connect with the vehicle networking management platform or cloud platform for efficient data transmission and sharing. The Uu interface serves as a communication bridge among the VIU, RSU, and base station, guaranteeing reliable communication with the mobile network and stable data transmission support for the intelligent transportation system. The PC5 interface represents the direct communication link between VIUs and between VIUs and RSU.

**Table 1. The Comparison of Performance Between the Two Communication Technology Standards**

| Performance indicators | DSRC | C-V2X |
|---|---|---|
| **Application scenarios** | Shorter distances, suitable for close-range communication between vehicles | Wider distance coverage, suitable for communication between vehicles and between vehicles and infrastructure |
| **High-density transmission support** | Packet loss occurs | Ensure no packet loss |
| **Reliability** | In some cases, it may not be stable enough and is susceptible to interference. | High reliability, especially in high-speed mobile environments. |

As shown in Table 1, from a technical performance perspective, C-V2X demonstrates significant advantages over DSRC in several key indicators, including its outstanding capacity performance, extremely low latency characteristics, and excellent manageability. Optimized for high-speed mobility, C-V2X surpasses DSRC in technical performance, particularly in link budget and communication range, which can double or offer greater reliability within the same span. Moreover, it provides robust support for high-speed vehicles and wide-ranging notifications, ensuring seamless and stable communication even in the presence of obstacles and blind spots.

In addition, many scholars have been conducting simulation experiments and real-world experiments to compare the performance of these two communication technologies. For example, Ghodhbane et al. used simulations to compare the performance of C-V2X and DSRC. The results showed that although the increase in network load reduced the performance of C-V2X, it outperformed DSRC in various indicators, including data packet intervals and packet reception rates. Maglogiannis et al. compared the performance of C-V2X and DSRC in a highway scenario. The experiment was conducted on an 8 km highway in the Netherlands using seven RSUs and two Vehicle Intelligent Units. V2X messages were first stored locally and then collected into a database by a central log server. The results showed that the communication range of C-V2X was higher than that of DSRC.

Therefore, in summary, C-V2X can provide better services for vehicle-road cooperative systems compared to DSRC. Based on the latest C-V2X communication technology standards, the proposed mechanism for information sharing among nearby vehicles in this paper demonstrates stronger feasibility and reliability.

## THE KEY TECHNOLOGIES FOR IMPLEMENTING THE SHARING MECHANISM

### The deployment method of the Road Section Management Units

When deploying Road Section Management Units, two common methods are typically used: hotspot area deployment and demand-driven deployment. Hotspot deployment usually refers to segments with complex traffic flows and high density, while demand-driven deployment refers to road sections where accidents occur frequently, with special traffic scenarios or prone to traffic congestion.

On highways, RSMU typically have extensive communication capabilities, and their communication range can usually cover distances of up to approximately 1 kilometer. However, in areas approaching the communication coverage limit, signal fluctuations may occur. This phenomenon could be related to complex environmental changes and the performance of communication equipment. Especially in scenarios like highways, when approaching the coverage limit, a noticeable inflection point in the packet loss rate can be observed, indicating that the relationship between packet loss rate and distance is not a simple linear change. The maximum effective connection for V2I communication is typically within a range of about 600 meters. Therefore, when considering the coverage quality and economic practicality of Road Section Management Units, the distance between two units should be controlled to approximately 1.2 kilometers. As for the installation height of Road Section Management Units, it should be ensured that there are no obstructions.

The real-time sharing mechanism of adjacent area vehicle driving information proposed in this paper is currently only discussed in the context of highways. Based on the direction of vehicle movement on highways, the deployment scenarios of Road Section Management Units can be roughly divided into three situations: the mainline of the highway, the exits of complex traffic flow ramps, and the entrances of ramps. The following sections will provide detailed analysis of these three scenarios.

The mainline of the highway：For the mainline of the highway, traffic generally flows in two directions, and the traffic volume is usually high. To facilitate the management of Road Section Management Units, units are installed on both sides of the lanes to manage vehicles traveling in different directions, as shown in Figure 2. The distance between two Road Section Management Units is approximately 1.2 kilometers, and the deployment height ensures unobstructed visibility.

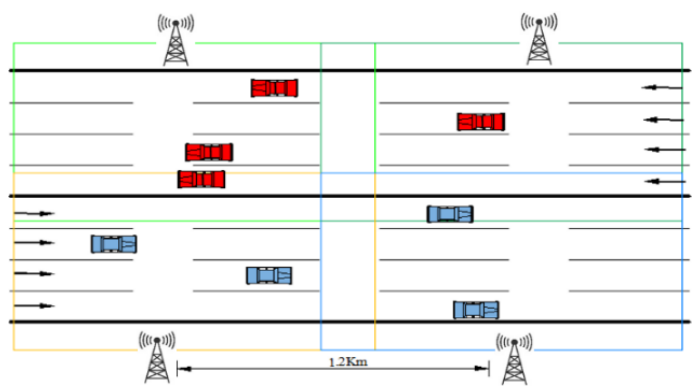

**Figure 2. Deployment method of Road Section Management Units on the mainline of highways.**

Complex traffic flow at highway ramp exits：For complex traffic flow at highway ramp exits, where there are significant variations in vehicle driving status and trajectories in the vicinity before exiting, this area poses significant safety hazards for vehicles. To address this, Road Section Management Units nodes can be deployed at specific points, such as 100 meters before the exit point and 50 meters after the exit point, as illustrated in Figure 3. The deployment height should be above 10 meters.

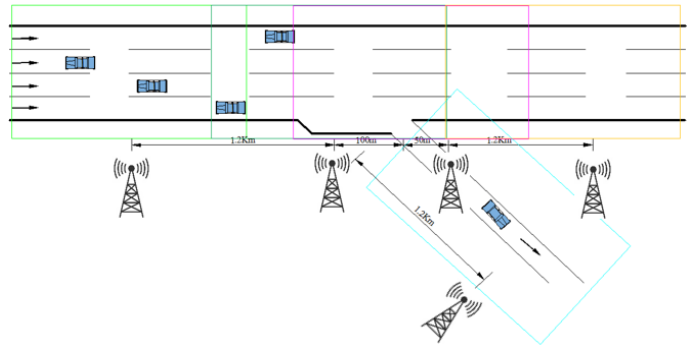

**Figure 3. Deployment method of Road Section Management Units at highway ramp exits.**

Highway ramp entrances：At the entrance of highway ramps, traffic flow is often dense, leading to limited visibility and significant blind spots in vehicle perception, posing a potential threat to driving safety. Within a certain distance after vehicles merge, there are considerable variations in driving status and trajectories, presenting a serious safety threat and substantial risk. Road Section Management Units nodes can be deployed approximately 100 meters before the merging point, as depicted in Figure 4. The height of the Road Section Management Units should also be above 10 meters.

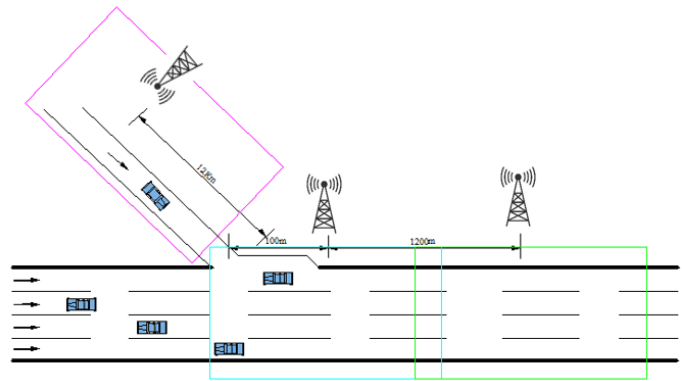

**Figure 4. Deployment method of Road Section Management Units at highway ramp entrances.**

### Mechanism of implementation for shared mechanism

The real-time information sharing mechanism among vehicles in adjacent areas proposed in this paper is based on ensuring time synchronization between Vehicle Intelligent Units and Road Section Management Units. It utilizes The basic framework real-time information sharing between Vehicle Intelligent Units and Road Section Management Units, as well as between different Road Section Management Units, to achieve real-time information sharing among vehicles in adjacent areas. The framework diagram of the information sharing mechanism is shown in Figure 5.

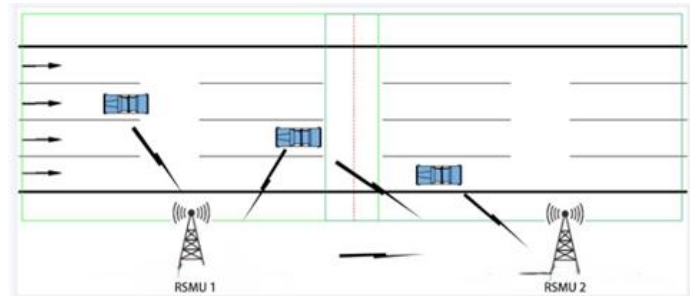

**Figure 5. Framework diagram of information sharing mechanism.**

As shown in Figure 5., in the real-time information sharing mechanism, the timing of the communication link between the Vehicle Intelligent Units and the Road Section Management Units is crucial. When the vehicle's Vehicle Intelligent Units 1 approaches the jurisdictional boundary of Road Section Management Units 1, it begins to establish an exclusive communication link with Road Section Management Units 1, transmitting only the vehicle's operational status and future driving intentions to it. As the Vehicle Intelligent Units 1 continues to travel, when it is about to leave the jurisdiction of Road Section Management Units 1 and enter the jurisdiction of Road Section Management Units 2, to ensure the continuity and stability of communication, Vehicle Intelligent Units 1 simultaneously establishes communication links with both Road Section Management Units 1 and Road Section Management Units 2. This parallel communication method not only prepares for subsequent communication with Road Section Management Units 2 but also ensures that there is no loss or delay of information when switching communication objects. However, at this stage, the information from Vehicle Intelligent Units 1 is still managed by Road Section Management Units 1 to ensure the unity and accuracy of information. When Vehicle Intelligent Units 1 completely leaves the jurisdiction of Road Section Management Units 1 and enters the jurisdiction of Road Section Management Units 2, it automatically disconnects from Road Section Management Units 1 and establishes a new exclusive communication link with Road Section Management Units 2. At this point, all driving information of Vehicle Intelligent Units 1 will be managed by Road Section Management Units 2, thereby achieving seamless communication. Adjacent Road Section Management Units also need to share information between them.

The specific process of establishing communication connection between the Vehicle Intelligent Units and each Road Section Management Units is as follows:

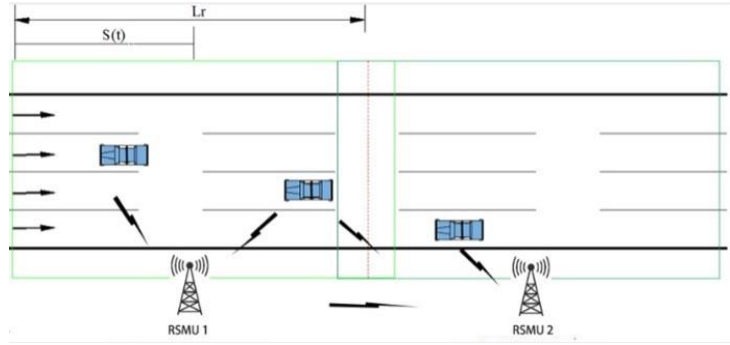

**Figure 6. Information sharing communication diagram.**

$L_r$ in Figure 6. represents the jurisdictional scope of the Road Section Management Units, while $S(t)$ represents the distance traveled by the Vehicle Intelligent Units within the jurisdictional scope of the Road Section Management Units.

**CONCLUSION**

This paper focuses on the application of the real-time information sharing mechanism between vehicles in the neighboring region and the spatio-temporal global view construction method for the operation period of highway transportation infrastructure in the highway scenario.

Establishment of a real-time information sharing foundation comprising a Road Section Management Unit and vehicle intelligent units. The Road Section Management Unit maintains real-time information on current vehicle trajectories, future driving intentions, road infrastructure, and abnormal events within their jurisdiction, sharing this information with all vehicles in their area and the adjacent Road Section Management Unit. Vehicle Intelligent Units upload vehicle attributes, driving intentions, and driving status information to the Road Section Management Unit and receive shared information from them.

Proposing a seamless distributed real-time information sharing mechanism based on the Road Section Management Unit. Each road section management unit uploads information about its managed sections to the cloud, and vehicle intelligent units' pre-download this information and directly establish links with the management unit upon entering any segment, reducing the cost of establishing data links. All Road Section Management Units share information about vehicle driving in their managed segments and the status of road infrastructure with neighboring Road Section Management Units, realizing a seamless distributed real-time information sharing mechanism.

Although the current research focuses on highway scenarios, it has laid a solid foundation for application scenarios such as other grades of highways and urban roads. In the future, with the continuous progress of technology and in-depth expansion of research, it is expected that this result will be widely promoted to further optimize the traffic information flow of all types of roads.

**REFERENCES**

Ahmed, E., Gharavi, H. (2018). "Cooperative Vehicular Networking: A Survey." IEEE Transactions on Intelligent Transportation Systems, 19(3): 996-1014.

Ghodhbane, C., Kassab, M., Maaloul, S., et al. (2022). "A Study of LTE-V2X Mode 4 Performances in a Multiapplication Context." IEEE Access, 10: 63579-63591.

Chen, W.T, He, Z.C, Zhu, Y.T, et al (2023). "Signal-Vehicle Coordination Control Modeling and Roadside Unit Deployment Evaluation under V2I Communication Environment." Journal of Advanced Transportation, 2023.

Ghosh, S., Misra, I. S., Chakraborty, T. (2023). "Optimal RSU deployment using complex network analysis for traffic prediction in VANET ." Peer-to-Peer Networking and Applications, 16(2): 1135-1154.

GLOBAL W. (2023). "Report on Road Safety 2023." . World Health Organization：Geneva，Switzerland，2023, Available online: https：//www.who.int/teams/social-determinants-of-health/safety-and-mobility/global-status-report-on-road-safety-2023.

Guanxu, Y. (2023). "Research on Communication Performance Analysis Algorithm of Large-scale C-V2X Equipment and Design and Development of Test System." JILIN UNIVERSITY, 2023.

Hu, Z., Zheng, Z., Wang, T., et al. (2017). "Roadside Unit Caching: Auction-Based Storage Allocation for Multiple Content Providers0." IEEE Transactions on Wireless Communications, 16(10): 6321-6334.

Hou, Y., Zhang, Z., Li, W. Z., et al. (2023). "Digital Twin based Packet Reception Prediction for C-V2X Networks." 98th IEEE Vehicular Technology Conference (VTC-Fall). Hong Kong,2023.

Jameel, F., Javed, M. A., Ngo, D. T. (2020). "Performance Analysis of Cooperative V2V and V2I Communications Under Correlated Fading." IEEE Transactions on Intelligent Transportation Systems, 21(8): 3476-3484.

Ko, B., Liu, K., Son, S. H., et al. (2021). "RSU-Assisted Adaptive Scheduling for Vehicle-to-Vehicle Data Sharing in Bidirectional Road Scenarios." IEEE Transactions on Intelligent Transportation Systems, 22(2): 977-989.

Maglogiannis, V., Naudts, D., Hadiwardoyo, S., et al. (2022). "Experimental V2X Evaluation for C-V2X and ITS-G5 Technologies in a Real-Life Highway Environment." IEEE Transactions on Network and Service Management, 19(2): 1521-1538.

Yi, H., Lei, H. (2023) "Data transmission strategy based on transmission loss function in C-V2X network." Telecommunications Science, 39(12): 65-75.

Yao, W. B., Liu, J. Y., Wang, C., et al. (2023). "Learning-based RSU Placement for C-V2X with Uncertain Traffic Density and Task Demand." IEEE Wireless Communications and Networking Conference (WCNC). Glasgow, SCOTLAND.

Zhang, L. Y., Wang, L., Zhang, L. L., et al. (2023). "An RSU Deployment Scheme for Vehicle-Infrastructure Cooperated Autonomous Driving." Sustainability, 15(4).